\documentclass{PoS}

\title{Top Quark Pair Production Cross Section at the Tevatron}

\ShortTitle{Top Quark Pair Production Cross Section at the Tevatron}

\author{\speaker{Reinhild Yvonne Peters}\thanks{on behalf of the CDF and D0 Collaborations.}\\
        The University of Manchester, also at DESY\\
         School of Physics and Astronomy,
         Oxford Road \\
         Manchester M13 9PL \\
         England\\
             E-mail: \email{repeters@cern.ch}}

\abstract{The top quark, discovered in 1995 by the CDF and D0
  collaborations at the Tevatron proton antiproton collider at
  Fermilab, has undergone intense studies in the last 20 years. Currently,
  CDF and D0 converge on their measurements of top-antitop quark
  production cross sections using the full Tevatron data sample. In
  these proceedings, the latest results on inclusive and differential
  measurements of top-antitop quark
  production cross sections at the Tevatron are reported. }

\FullConference{The European Physical Society Conference on High Energy Physics\\
		22--29 July 2015\\
		Vienna, Austria}

\begin{document}

\section{Introduction}
Discovered more than 20 years ago, in 1995, by the CDF and D0
collaborations~\cite{cdfobs,d0obs} at the Fermilab Tevatron proton
antiproton collider, the top quark is the heaviest elementary particle
known today. Its high mass might indicate that it plays a special role
in electroweak symmetry breaking and makes it a promising candidate as window
to new physics. The detailed study of the top quark and its production
mechanism is therefore crucial. 

In this article, measurements of top-antitop quark
production ($t\bar{t}$) cross sections, performed by CDF and D0 in
Run~II of the Tevatron, are discussed. In Run~II, lasting from 2001 to
2011, the collision energy
was 1.96~TeV. The measurements
of inclusive $t\bar{t}$ cross sections and their comparison to
predictions from quantum chromodynamics (QCD) calculations allows to
test the standard model (SM) in detail. Many new physics models with final
states similar to that of $t\bar{t}$ production would change the
measured cross sections significantly. Furthermore, differential
measurements of $t\bar{t}$ cross sections are reported, which serve as
test for perturbative QCD (pQCD) predictions, can be used to tune
Monte Carlo simulations, and are a useful tool to look for new physics
that would change the shape of kinematic or topological variables
relative to the SM prediction.


\section{Inclusive Cross Section Measurements}
The production of $t\bar{t}$ pairs occurs via the strong
interaction. At the Tevatron, approximately 85\% of the $t\bar{t}$
production happen via quark-antiquark annihilation, and 15\% via
gluon-gluon fusion. These proportions are approximately inverted at
the LHC. This means that the sensitivity to beyond the SM (BSM)
physics scenarios is different at Tevatron and LHC.

The measurement of the $t\bar{t}$ cross section is done separately in
different final states. In the SM, the top quark decays almost 100\% of the time
into a $W$ boson and a $b$~quark, with the $W$ either decaying into a
quark-antiquark pair, or a charged lepton and the corresponding
neutrino. The event selection in each decay channel is adjusted to the
respective final state objects, with the goal to enhance the data sample with
$t\bar{t}$ events and reduce the background, while keeping a
reasonable statistics. The decay channels considered are usually
the dilepton final state, where both $W$~bosons decay into electrons,
muons or leptonically decaying taus,
the semileptonic final state, with one $W$~boson decaying into
electron, muon or leptonically decaying tau, 
and the other $W$ boson into quarks, and the fully hadronic final state, in
which both $W$ bosons decay into quarks. In the dilepton final state a
low background can be achieved, but the branching fraction is small,
while allhadronic events represent a large branching fraction, but a
large background. The best mixture for most analyses
is usually the semileptonic channel, which mixes a good branching
fraction with a manageable background. Depending on the channel,
different approaches to the measurement of the cross section are
done. In very clean channels usually a counting-method is sufficient,
while less clean ones require the use of multivariate analysis
techniques, in which several variables with small signal-to-background
separation are combined into one discriminant. Given the two $b$-jets
in the $t\bar{t}$ final state, a useful tool often applied
is $b$-tagging, where properties of displaced tracks and secondary
vertices from the $B$~hadron decay are considered to identify jets
from $b$~fragmentation.

In this article, analyses in these three main final states are
presented. The main background contribution in dileptonic events comes
from $Z$+jets processes, which can be simulated using Monte Carlo (MC)
generators. In semileptonic events, the handling of the largest
background contribution, $W$+jets, is done using simulated events and
normalizing the yield using a data-driven approach. For multijet
events, the largest contribution comes from QCD multijet events, which
require data-driven modelling of the background. Further, smaller
background contributions are single top, diboson and fake events,
where the latter are events coming from jets misidentified as leptons
in the detector. 

\subsection{Tevatron Combination}
In 2014, CDF and D0 published the first combination of the $t\bar{t}$
cross section from both collaborations, using data samples between 2.9
and 8.8~fb$^{-1}$. In particular, four analyses from CDF and two from
D0 were used in the combination. CDF provided one analysis in the
dilepton final state, where the number of events with at least one
$b$-tagged jet was counted, two analyses in semileptonic events,
where the number of events with at least one
$b$-tagged jet was counted or a Neural-Network discriminant~\cite{tmva} was
built based on kinematic variables. The fourth analysis, in the
allhadronic final state, used events with exactly one or at least one
$b$-tagged jet, performing a maximum likelihood fit to the
reconstructed top quark mass. 
The D0 collaboration provided an analysis in the dileptonic final
state, where a likelihood fit to a discriminant  based on a
Neural-Network $b$-tagging algorithm, was performed, and an analysis
in the semileptonic final state was performed, in which events with three and more
than three jets were split into events with zero, one or at least two
$b$-tagged jets. For signal dominated sub-channels a simple counting
method was then used, while in background-dominated sub-channels a
random forest discriminant was applied.  

\begin{figure*}[t]
\centering
\includegraphics[width=88mm]{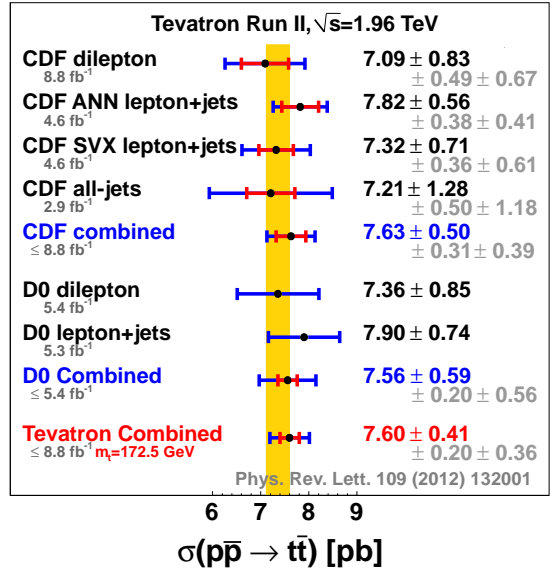}
\caption{ Tevatron combination of the $t\bar{t}$ cross section by the
  CDF and D0 collaborations~\cite{tevcombi}. } \label{fig:tevcombi}
\end{figure*}

The  CDF analyses were combined using BLUE~\cite{blue}, while the D0 analyses were fitted simultaneously with a likelihood fit,
where the systematic uncertainties were treated as nuisance
parameters. These two experiment-specific combinations were then
combined in a further step using BLUE, resulting in a $t\bar{t}$ cross
section of $\sigma_{t\bar{t}} = 7.60 \pm 0.41 {\rm
  (stat+syst)}$~pb~\cite{tevcombi} for a top quark mass of $172.5$~GeV. 
This result is compatible with the SM prediction of
$\sigma_{t\bar{t}}= 7.35^{+0.23}_{-0.27} {\rm (scale+pdf)}$~pb, calculated at
next-to-next-to-leading order  (NNLO) precision in QCD, with soft-gluon
resummation to 
next-to-next-to-leading logarithmic
accuracy~\cite{czakon}. The cross section measurements in
different channels are compatible with each other, showing no evidence
for possible BSM contributions. Figure~\ref{fig:tevcombi} shows the
$t\bar{t}$ cross sections in the different channels by the two
experiments, as well as the combination.


\subsection{New D0 result}
Using the full Tevatron data sample of 9.7~fb$^{-1}$, the D0
collaboration recently updated their inclusive measurement of the
$t\bar{t}$ production cross section in the dileptonic and semileptonic
final states. The analysis strategy is similar to the one used for the
Tevatron combination: in dileptonic events, the discriminant trained
for $b$-jet identification via a  multivariate analysis technique is
used to discriminate $t\bar{t}$ signal from the background. A template
fit is performed to extract the cross section value. In semileptonic
events, a topological discriminant, based on boosted decision trees
(BDTs), is trained separately for events with two, three or at least
four jets, and the cross section is extracted by simultaneously
performing a maximum likelihood fit on the  discriminants in these
three sub-channels. The resulting cross section is  $\sigma_{t\bar{t}} = 7.73 \pm 0.13 {\rm
  (stat)} \pm 0.55 {\rm (syst)}$~pb~\cite{d0inclxsec}, in good
agreement with the SM prediction. The main contribution to the
systematic uncertainties comes from modelling of the hadronization.

\section{Differential Cross Section Measurements}
Besides performing measurements of inclusive cross sections, the
determination of the $t\bar{t}$ cross section differentially as
function of various variables is important. Differential $t\bar{t}$
cross section measurements provide a test of pQCD, help to tune the
simulation of $t\bar{t}$ events, and can yield
additional insight into potential hints for new physics. For example,
deviations in the $t\bar{t}$ cross section as function of the
invariant $t\bar{t}$ mass, $m_{t\bar{t}}$, could yield insight into
the existence of heavy resonances decaying into a pair of top quarks,
or possible new physics contributions to the $t\bar{t}$
forward-backward asymmetry can be probed by measuring differential
distributions in variables related to the pseudorapidity of the top
quark. 

Using the full Run~II data sample of 9.7~fb$^{-1}$, D0 recently
performed a measurement of the $t\bar{t}$ cross section as function of
three variables: $m_{t\bar{t}}$, the absolute value of the rapidity of
the top quark, $|y^{top}|$, and the transverse momentum of the top
quark, $p_T^{top}$~\cite{diffxsec}. The analysis was performed in the semileptonic
final state, with at least one jet required to be $b$-tagged. 
The calculation of the variables  $|y^{top}|$ and $p_T^{top}$ requires
the assignment of final state objects to originate from the top or the
antitop quark. For this assignment, a constrained kinematic reconstruction
algorithm~\cite{recoalgo} is used, in which experimental resolutions are taken
into account. In this algorithm, the top quark mass and $W$~boson mass
are fixed to their known values, allowing to determine the
$z$-component of the neutrino momentum from the $W$ boson mass
constraint and reducing the number of possible jet-quark assignments
via the top quark mass constraint. The solution with the best $\chi^2$
is taken for the construction of the top quark vectors. 

\begin{figure*}[t]
\centering
\includegraphics[width=120mm]{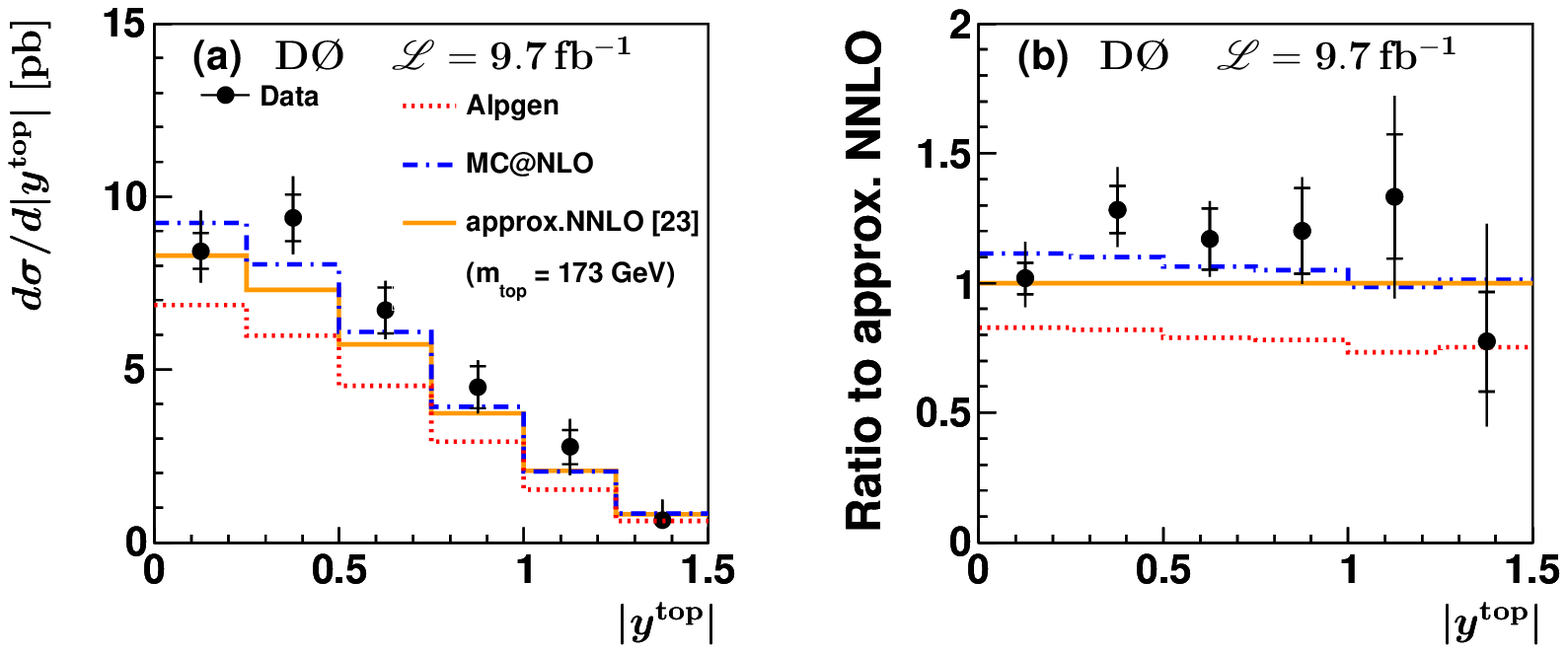}
\includegraphics[width=120mm]{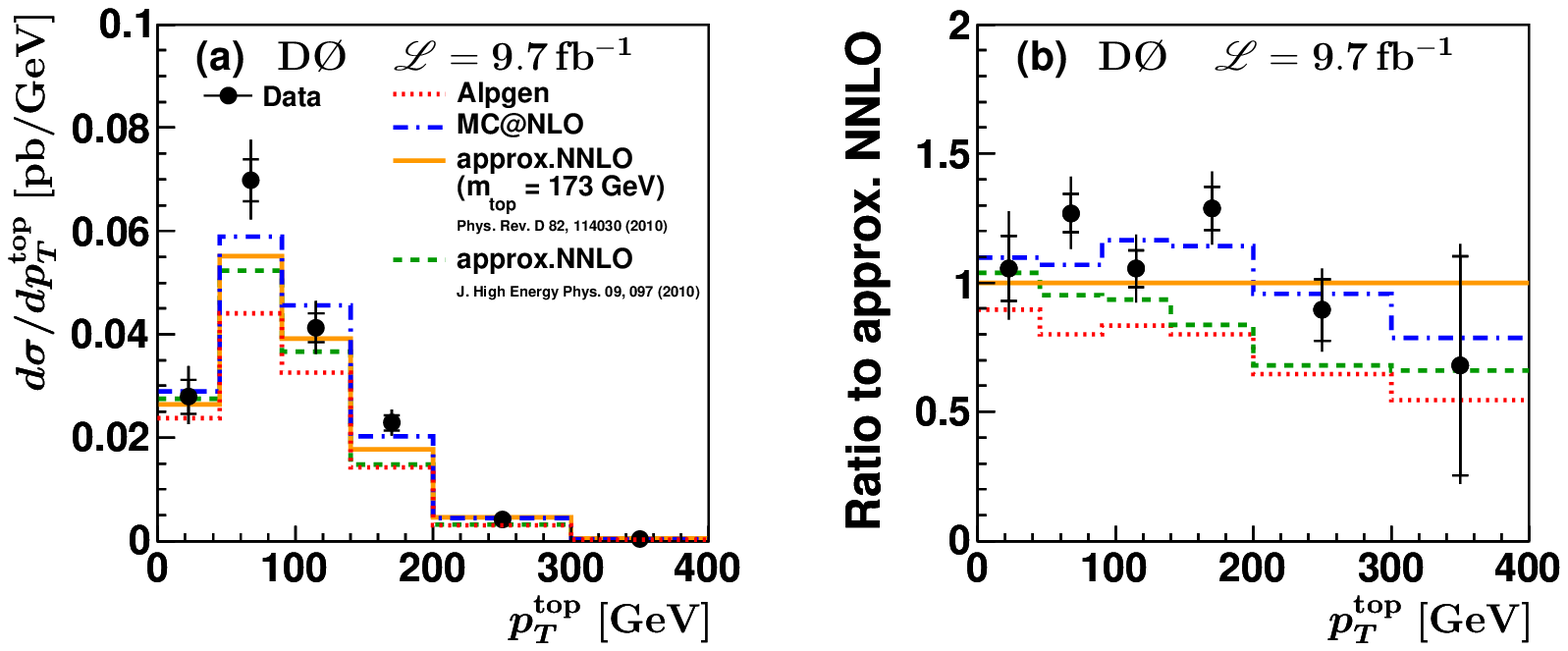}
\caption{ The $t\bar{t}$ cross section measured as function of
  $|y^{top}|$ (top), and $p_T^{top}$
  (bottom). The left figures show the cross section, the right figures
  the ratio of the measured cross section with respect to the
  approximate NNLO calculation~\cite{diffxsec}. } \label{fig:differential}
\end{figure*}

The measurement is defined for parton-level top quarks, including
off-shell effects. Therefore, a correction for detector and acceptance
effects has to be done. A regularized unfolding, implemented in
{\verb{TUNFOLD}}~\cite{tunfold} is used for this purpose, where the regularization is based on
the derivative of the distributions. To keep as much information as
possible, the number of bins for the reconstructed distributions is
kept twice as high as the number of bins on parton level. The
contribution of different sources of systematic uncertainties are
evaluated by changing the migration matrix and the background
contribution. The largest uncertainties arise for high bins in
$m_{t\bar{t}}$,  $|y^{top}|$, and $p_T^{top}$. 
In Figure~\ref{fig:differential} the measured $t\bar{t}$ cross section
differentially in  $|y^{top}|$, and $p_T^{top}$ is
shown, together with predictions from Monte Carlo simulation and
predictions from approximate NNLO calculations. In general, the
agreement is good relative to the generator and approximate NNLO
predictions. For ALPGEN~\cite{alpgen}, the absolute normalization is
too low, while in the  $|y^{top}|$ variable the description by the
MC@NLO generator~\cite{mcnlo} is better than that of the approximate
NNLO prediction. 

\begin{figure*}[t]
\centering
\includegraphics[width=120mm]{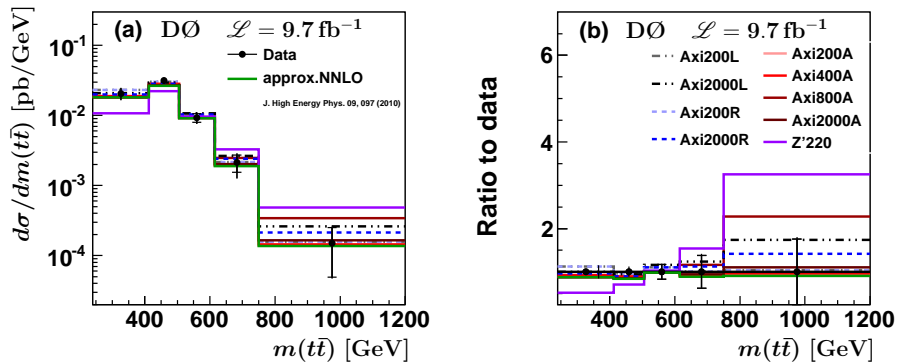}
\caption{ The $t\bar{t}$ cross section measured as function of
  $m_{t\bar{t}}$ and compared to various axigluon
  models~\cite{diffxsec}. } \label{fig:axigluons}
\end{figure*}

The $t\bar{t}$ forward-backward asymmetry has been measured by both
CDF and D0 in recent years to be  somewhat larger than 
predicted by the SM~\cite{asymreview}. In many BSM models, for example
$Z^{'}$ or axigluon models,  this asymmetry would be associated to a
change in  differential distributions.  Figure~\ref{fig:axigluons} shows the differential $t\bar{t}$
cross section as function of $m_{t\bar{t}}$, compared
to different axigluon and a $Z^{'}$ model. The measurement excludes
several of these models.

\section{Conclusion and Outlook}
The measurements of the inclusive and differential $t\bar{t}$ cross
sections using the full data samples collected by the CDF and D0
collaborations at the Tevatron is moving towards its completion. These
legacy measurements are important to test the SM, tune Monte Carlo
simulations and search for hints of new physics. The different initial
state and energy of the Tevatron compared to 
the LHC,  make these complementary measurements.

\section*{Acknowledgments}
I thank my collaborators from CDF and  D0
 for their help in preparing the presentation and this
article. I also thank the staffs at Fermilab and
collaborating institutions, and acknowledge the support from the
Helmholtz association.

\end{document}